\documentclass[%
 reprint,
 longbibliography,
 amsmath,amssymb,
 aps,
pra,
]{revtex4-1}
\usepackage{graphicx}
\usepackage{dcolumn}
\usepackage{bm}
\newcommand{\ud}{\mathrm{d}}

\begin{document}

\preprint{APS/123-QED}

\title{Evolution of the Hofstadter butterfly in a tunable optical lattice}

\author{F. Y{\i}lmaz}
\email{firat.yilmaz@bilkent.edu.tr}
\author{F. Nur \"{U}nal}%
\author{M. \"{O}. Oktel}
\affiliation{Department of Physics, Bilkent University, Ankara, Turkey}




\date{\today}

\begin{abstract}
Recent advances in realizing artificial gauge fields on optical lattices promise experimental detection of topologically non-trivial energy spectra. Self-similar fractal energy structures generally known as Hofstadter butterflies depend sensitively on the geometry of the underlying lattice, as well as the applied magnetic field. The recent demonstration of an adjustable lattice geometry [L. Tarruell \textit{et al.}, Nature 483, 302--305 (2012)] presents a unique opportunity to study this dependence. In this paper, we calculate the Hofstadter butterflies that can be obtained in such an adjustable lattice and find three qualitatively different regimes. We show that the existence of Dirac points at zero magnetic field does not imply the topological equivalence of spectra at finite field. As the real-space structure evolves from the checkerboard lattice to the honeycomb lattice, two square lattice Hofstadter butterflies merge to form a honeycomb lattice butterfly. This merging is topologically non-trivial, as it is accomplished by sequential closings of gaps. Ensuing Chern number transfer between the bands can be probed with the adjustable lattice experiments. We also calculate the Chern numbers of the gaps for qualitatively different spectra and discuss the evolution of topological properties with underlying lattice geometry.
\begin{description}
\item[PACS numbers]
\end{description}
\end{abstract}

\pacs{Valid PACS appear here}
\maketitle


\section{Introduction}
The self-similar behavior of the energy spectrum of a charged particle moving in a lattice under a magnetic field has attracted great interest since its first prediction in 1976 \cite{Hofstadter}. Such energy spectra are called Hofstadter butterflies due to the visually pleasing symmetry of the original square lattice calculation. However, experimental observation of the Hofstadter butterfly requires magnetic fields thousands of times stronger than what is currently attainable for conventional solid state systems. In a usual solid state lattice, the length scale of the periodic potential is much smaller than the magnetic length. This obstacle has steered the efforts towards the direction of artificial regulation of lattice parameters or creation of synthetic magnetic fields.

First attempts to experimentally realize the Hofstadter butterfly used superlattices which aim to enhance the lattice scale up to magnetic length \cite{SuperlatticeSchweizer,SuperlatticeRegister,SuperlatticeKim}. These systems lack the adjustability of the lattice geometry, constituent particles and interactions which are all possible with a cold atom system in a defect-free environment. One of the major accomplishments in cold atom experiments is the creation of effective magnetic fields through laser assisted tunneling \cite{BlochHofstadter,KetterleHarper}. These experiments are taking firm steps towards realizing degenerate Fermi and Bose gases with self-similar energy spectra.

Another exciting development is the demonstration of an adjustable optical lattice which was used to investigate the dynamics of Dirac points \cite{EsslingerDirac,EsslingerGraphene,EsslingerDoubleTransfer,EsslingerQuantumMagnetism}. This lattice can be adiabatically changed from a dimer to a honeycomb geometry. An artificial gauge field has recently been added \cite{EsslingerHaldane} to the same setup to realize the Haldane model \cite{Haldane}. It is thus reasonable to expect a constant uniform magnetic field to be simulated in this lattice. An adjustable optical lattice with a uniform artificial magnetic field would make it possible to investigate not only the energy spectrum but also the relation between the lattice geometry and topological properties.

In this paper, we study the Hofstadter butterflies and their topological nature for lattice geometries that can be created by the Zurich group. We show that this tunable optical lattice captures qualitatively different spectra and enables the study of merging of two square lattices butterflies to form the spectrum of a honeycomb lattice. This merging is topologically non-trivial as an infinite number of gaps in the spectrum must close during this transition.

We numerically solve the Schr\"{o}dinger equation for the potentials reported in Ref.\cite{EsslingerDirac,EsslingerGraphene} and obtain the lowest two bands. In the parameter regime we are interested in, these bands can be accurately described by a nearest neighbor tight-binding approximation. We find the hopping strengths by fitting band structure over the whole Brillouin zone. Magnetic field is introduced to the tight-binding Hamiltonian through Peierls substitution. If the magnetic flux per unit cell of the lattice (in units of flux quantum $\phi_0=h/e$) is a rational number $\Phi=p/q$, energy spectra can be obtained efficiently by diagonalizing a $q\times q$ matrix. We calculated the Hofstadter butterfly spectra for all the different regimes that can be accessed by this adjustable optical lattice.

In the previous studies concerning adjustable or deformed real-space lattices, the presence of Dirac points in the spectra has been used to mark qualitative change in the physics of the system. For example, a lattice that lacks six-fold symmetry was successfully used to mimic graphene \cite{EsslingerGraphene}. While it is natural to expect the low energy physics to be dominated by the Dirac cones if the Fermi surface lies close to them, it is not clear if the presence of Dirac cones qualitatively determine the global shape of the energy spectrum. In this sense, one can ask whether the presence of Dirac cones at zero magnetic field for two different systems necessitates their topological equivalence at non-zero magnetic field. We show that this equivalence is not guaranteed, \textit{i.e.} two lattices both of which have Dirac points at zero field may have qualitatively different behavior at non-zero field and may have different topological characters. More specifically, as the lattice alters from the checkerboard to the honeycomb geometry, the energy spectrum changes from two separate square lattice butterflies to the honeycomb lattice butterfly. However, the energy gap around zero closes at different points during the transition depending on the denominator $q$ of the flux $\Phi=p/q$. We characterize the topological nature of the transition by calculating the Chern numbers for the major gaps of the spectrum.

The paper is organized as follows: In the next section, we summarize the properties of the lattice generated by the Zurich group and outline our numerical procedure for the tight-binding description. The third section, we discuss the effect of the magnetic field and solve the resulting Hamiltonian to obtain the Hofstadter butterfly in the interesting parameter regime. Sec. IV presents the calculation of the first Chern numbers in the butterfly and a detailed discussion of the transition. We conclude in Sec. V with a summary and experimental consequences of our results.

\section{THE MODEL}
We start by considering the single-particle spectrum in a two-dimensional periodic potential in the absence of the magnetic field. As long as the Wannier functions for the lowest two bands are well localized, the effect of the magnetic field can be included by modifying the hopping parameters via the Peierls substitution \cite{Peierls}. To obtain the tight-binding parameters, we first calculate the energy spectrum of the system by numerical solution of the Schr\"{o}dinger equation. The real-space Hamiltonian without the magnetic field is
\begin{equation}
{\cal H}=-\frac{\hbar^2}{2m}\bigg(\frac{\partial^2}{\partial x^2}+\frac{\partial^2}{\partial y^2}\bigg)+ V(x,y), \label{Hamiltonian}
\end{equation}
where we use the potential $V(x,y)$ reported by the Zurich group \cite{EsslingerDirac},
\begin{align}
V(x,y)=&-V_{\bar{x}}\cos^2(kx+\frac{\theta}{2})-V_{x}\cos^2(kx)-V_{y}\cos^2(ky)\nonumber\\
&-2\alpha\sqrt{V_xV_y}\cos(kx)\cos(ky)\cos(\varphi),
\end{align}
with $k=2\pi/\lambda$, $\lambda$ being the wavelength of laser beams creating the optical lattice. $\alpha$ is the visibility of the interference pattern. The phases $\theta=\pi$ and $\varphi=0$ are set to the experimental values. The potential depths we use are in the experimentally reported regime. The confinement along the $z$-direction is strong so that the system consists of independent layers of two-dimensional sheets.

As the parameters $V_{\bar{x}}, V_x$ and $V_y$ are varied, the Bravais lattice defined by the periodicity of the potential does not change. However, within a unit cell, the number and positions of potential minima vary determining the overall lattice geometry. If the potentials are deep enough compared to recoil energy $E_R=\frac{h^2}{2m\lambda^2}$, states localized around these minima dominate the low energy physics and determine the structure of the lowest bands in the spectrum. The lattice formed by these states span a range of geometries varying from the checkerboard to the honeycomb lattice. Within this range, the most important qualitative change is the creation of Dirac points in $k$-space, which was investigated in detail both experimentally and theoretically \cite{EsslingerDirac,EsslingerGraphene,MontambauxPRL}. At zero magnetic field, we follow a similar procedure to obtain a tight-binding description.

We rotate the coordinate system by $\pi/4$ and then use $\lambda/\sqrt{2}$ to make positions $x,y$ dimensionless making the unit cell a one-by-one square in the $x-y$ plane. All the potentials and energies are measured in units of recoil energy $E_R$. We obtain numerical solutions of this dimensionless Hamiltonian for the lattice parameters defined above. We discretize the real-space Hamiltonian Eq.\ref{Hamiltonian} and numerically diagonalize the resulting energy matrix for 2500 points per unit cell with boundary conditions determined by the Bloch theorem.

\begin{figure}
\includegraphics[width=0.47\textwidth]{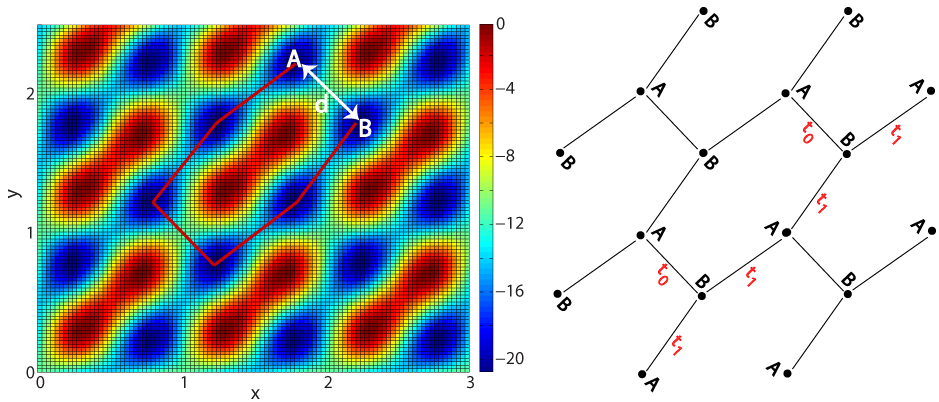}
\caption{(Color online) The real-space potential and corresponding tight-binding lattice. The unit cell remains square for all experimental parameters \cite{EsslingerGraphene}. The tunneling amplitudes $t_0,t_1$ are calculated by a fit to numerically obtained two lowest band energies. For $V_{\bar{x}}=14,V_x=0.79,V_y=6.45$ in units of $E_R$ and $\alpha=0.9$, we find $t_0 =0.0391,\:t_1 = 0.0361$. Within a unit cell, two well localized Wannier functions at potential minima (A,B) form the basis for tight-binding treatment. Modification of the potential depth parameters changes the distance $d$ between the sites A and B altering the tunneling rates between these sites. } \label{lattice Fig}
\end{figure}

For the parameter regime we are interested in, the lattice potentials are chosen deep enough to create narrow and symmetric energy bands, as the corresponding Wannier functions are well localized in real-space. Thus, a bipartite symmetric tight-binding approximation to these bands successfully describes the system. We construct the lattice with Bravais lattice vectors $\vec{R}_{m_1,m_2} = m_1 \vec{a_1}+ m_2\vec{a_2}$ for $(m_1,m_2)\in\mathbb{Z}$ for dimensionless $\vec{a_1}=\hat{x},\vec{a_2}=\hat{y}$. There can be one or two potential minima within the unit cell depending on the parameters. As we are interested in the emergence of Dirac points which corresponds to the transition from the dimer to the honeycomb-like lattice, we choose a unit cell with a two-state basis (see Fig.\ref{lattice Fig}). These states, labeled as $A$ and $B$, are localized at
\begin{equation}
\vec{d}_{A} = \frac{d}{2 \sqrt{2}} ( -\hat{x} + \hat{y} ), \quad \vec{d}_B = \frac{d}{2 \sqrt{2}} ( \hat{x} - \hat{y} ),
\end{equation}
where $d$ is the distance between the sites $A$ and $B$ in units of the lattice constant. In the experiments, adjusting the potential parameters modifies this distance $d$ which effectively changes the hopping parameter between sites $A$ and $B$. That is to say, without the magnetic field, the distance $d$ is embedded into tunneling amplitudes $t_0,t_1$ in the tight-binding Hamiltonian. As we work in the regime where the lowest two bands are narrow, we limit out tight-binding description to nearest-neighbor hopping. Each site has three nearest-neighbors; one in the same unit cell and two outside. Each state of type $A$ is connected to only states of type $B$. The hopping parameter $t_0$ connects the states within the unit cell while $t_1$ links neighboring unit cells. Due to the bipartite symmetry of the lattice, the lowest two bands are symmetric in energy.
The tight-binding Hamiltonian follows:
\begin{align}
{\cal H}=-\!\sum_{m_1,m_2}& \Big\{t_0|m_1  ,m_2  , A \rangle\langle m1, m2,B|\nonumber\\
&+t_1 | m_1+1,m_2  , A \rangle \langle m1, m2, B | \nonumber\\
&+ t_1 | m_1  ,m_2-1, A \rangle \langle m1, m2, B | + h.c. \Big\},
\end{align}
where $h.c.$ stands for Hermitian conjugate. The self energy term $\epsilon_0$ is excluded so that the energies are symmetric around zero.

The tight-binding Hamiltonian is readily diagonalized by Fourier transformation
\begin{align}
\epsilon &(t_0,t_1)=\nonumber\\
&\pm\sqrt{t_0^2+2t_1^2+2t_0t_1(\cos k_x+\cos k_y)+2t_1^2cos(k_x-k_y)},
\end{align}
where $k_x(k_y)$ is the quasi momentum along $x(y)$-direction. By fitting this expression to the numerical results (Eq.\ref{Hamiltonian}), we find the tunneling amplitudes $t_0,t_1$ together with the self energy $\epsilon_0$ for any optical lattice potential in an efficient way. In the earlier studies concerning the merging of Dirac points, the hopping parameters were determined by fitting these two energies specifically around the Dirac cones \cite{MontambauxPRL}. As we are interested in the global properties of the spectrum, we minimize the least-square error in energies of both bands over the entire Brillouin zone. For our calculations, this error is limited by $4.5\%$ as an average with respect to the band widths and reduces as we move towards the honeycomb geometry.

Therefore, the three parameters from the experiments (six parameters if the phases and the visibility are included) reduce to just two parameters in the tight-binding model $t_0,t_1$. When the energy is scaled by a tunneling rate, just a single parameter $t_1/t_0$ determines the spectrum. Each different lattice is represented by its $t_1/t_0$ ratio which gives a different butterfly spectrum when the magnetic field is applied.

As we only include the nearest-neighbor hoppings, the lattice is bipartite. In order to have a Dirac point, Schr\"{o}dinger equation must have a zero eigenvalue. For a state with zero energy the total incoming current to any lattice site must be zero. Each site has three neighbors all of which reside on the same sublattice. Thus, the wave function on each neighbor can vary by at most a phase factor. The existence of a zero energy solution depends on the solution of $t_0+t_1(e^{i\phi_1}+e^{i\phi_2})=0$. Therefore, we obtain the condition for the existence of Dirac points as $\frac{t_1}{t_0}\geq\frac{1}{2}$ at zero magnetic field in agreement with Ref.\cite{MontambauxPRL}. We stress that the existence of Dirac point at zero magnetic field does not imply the existence of zero energy solutions at finite field.

The effect of an external magnetic field can be included in the continuum Hamiltonian \ref{Hamiltonian} by minimal coupling $\vec{p}\rightarrow \vec{p}-e\vec{A}$. However, the modification of the lowest few bands by the magnetic field can be accurately described through tight-binding with complex hopping parameters. This procedure, called the Peierls substitution, was found to be precise as long as the tight-binding description at zero field holds \cite{OktelP}. For the recent cold atom experiments, this substitution method also captures the physical process \cite{BlochHofstadter,KetterleHarper}. The artificial magnetic fields in optical lattices are simulated by modification of the hopping between neighboring sites.

We use the Landau gauge $\vec{A} = (0,B_0 x,0)$ for the magnetic field $\vec{B}=B_0\hat{z}$. The tunneling amplitudes acquire a complex phase
\begin{equation}
t_{m,n}^{i,j} |n, j \rangle \langle m, i | \rightarrow e^{-i \Theta_{nm}^{ij}} t_{m,n}^{i,j} |n, j \rangle \langle m, i |,
\end{equation}
where $\Theta_{mn}^{ij} = \frac{e}{\hbar} \int_{(m,i)}^{(n,j)} \vec{A} \cdot \ud\vec{\ell}$, for lattice points $m=(m_1,m_2),\:n=(n_1,n_2)$ and sites $(i,j)$.
The hopping phase between two arbitrary lattice points is:
\begin{eqnarray*}
\Theta_{mn}^{ij} = 2 \pi \Phi
\left( \hat{y}\cdot ( \vec{R}_n-\vec{R}_m +\vec{d}_j - \vec{d}_i) \right)
\\
\times \left[ \hat{x} \cdot \frac{\vec{R}_n+\vec{R}_m +\vec{d}_j + \vec{d}_i) }{2}  \right].
\end{eqnarray*}
Therefore, the tight-binding Hamiltonian under the magnetic field becomes
\begin{align}
{\cal H} = -&\!\!\sum_{m_1,m_2} \!\Big\{t_0 | m_1  ,m_2,A \rangle\langle m_1, m_2,B| e^{-i 2\pi \Phi d' m_1} \nonumber\\
&+ t_1 | m_1+1,m_2,A \rangle\langle m_1, m_2,B| e^{-i 2 \pi \Phi d' (m_1 + \frac{1}{2})} \nonumber\\
&+ t_1 | m_1,m_2-1,A \rangle\langle m_1, m_2,B| e^{-i 2 \pi \Phi (d' - 1) m_1} + h.c.\Big\},\quad  \label{H_TB}
\end{align}
where $\Phi = \phi / \phi_0$ stands for the total magnetic flux $\phi$ passing through each unit cell over flux quantum $\phi_0 = h/e$ and $d' = \frac{d}{\sqrt{2}}$.

In the next section, we obtain the eigenvalues of this Hamiltonian and analyze the transition from the checkerboard to the honeycomb lattice.

\section{THE TIGHT-BINDING SOLUTION}
To find the eigenvalues, we consider rational flux values $\Phi=p/q$ where $p$ and $q$ are co-prime integers. Eigenstates of the Hamiltonian Eq.\ref{H_TB} can be expanded in real-space
\begin{equation}
|\Psi \rangle= \sum_{m_1,m_2} \psi^A_{m_1,m_2} |m_1,m_2,A\rangle + \psi^B_{m_1,m_2} |m_1,m_2,B\rangle.
\end{equation}
The Schr\"{o}dinger equation for the coefficients at sites A and B reads
\begin{align}
\epsilon  \psi^B_{m_1,m_2}=&-t_1 e^{i 2 \pi \Phi d' (m_1 + 1/2)} \psi^A_{m_1+1,m_2} \nonumber\\
&- t_0 e^{i 2 \pi \Phi d' m_1} \psi^A_{m_1,m_2}\nonumber\\
&-t_1 e^{-i 2 \pi \Phi (1-d') m_1} \psi^A_{m_1,m_2-1}, \label{S.E.A}
\end{align}
\begin{align}
\epsilon  \psi^A_{m_1,m_2} =&- t_1 e^{-i 2 \pi \Phi d' (m_1 - 1/2)} \psi^B_{m_1-1,m_2}\nonumber\\
&- t_0 e^{-i 2 \pi \Phi d' m_1} \psi^B_{m_1,m_2}\nonumber\\
&-t_1 e^{ i 2 \pi \Phi (1-d') m_1} \psi^B_{m_1,m_2+1}, \label{S.E.B}
\end{align}
where $\epsilon$ is the eigenvalue. These two sets can be combined to give a quadratic equation
\begin{align}
\epsilon ^2 \psi^A_{m_1,m_2} &=(t^2_0 + 2 t^2_1) \psi^A_{m_1,m_2}\nonumber\\
&+t_0 t_1 \Big\{ e^{i 2 \pi \Phi m_1}  \psi^A_{m_1,m_2-1} + e^{-i \pi \Phi d' } \psi^A_{m_1-1,m_2} \nonumber\\
&+ e^{i 2 \pi \Phi m_1} \psi^A_{m_1,m_2+1}+ e^{i \pi \Phi d' } \psi^A_{m_1+1,m_2} \Big\}\nonumber\\
&+ t_1^2\Big\{ e^{-i 2 \pi \Phi (m_1-1)} e^{-i \pi \Phi d'} \psi^A_{m_1-1,m_2-1}\nonumber\\
&+ e^{i 2 \pi \Phi m_1} e^{i \pi \Phi d' } \psi^A_{m_1+1,m_2+1}\Big\}. \label{Difference Eq.raw}
\end{align}
Since we employ the Landau gauge with the vector potential $\vec{A}$ along one of the primitive vectors of the lattice, we retain the translational symmetry along one direction. Thus, the $m_2$ dependent part of the wave function can be separated
\begin{equation}
\psi_{m_1,m_2}(k_y)= e^{i k_y m_2}e^{-i\pi\Phi d' m_1} f_{m_1}, \label{Psi_m1,m2(kx)}
\end{equation}
where $k_y$ is the quasi momentum along the $y$-direction. The second exponent term is introduced to make the eigenvalue equation independent of $d'$ through a gauge transformation. Reorganizing Eq.\ref{Difference Eq.raw} with the help of Eq.\ref{Psi_m1,m2(kx)} results in a one-dimensional difference equation,
\begin{align}
\lambda f_{m_1} &= \left( 1 + \frac {t_1}{t_0}e^{-i (2 \pi \Phi (m_1 -1) + k_y)} \right) f_{m_1-1} \nonumber\\
&+ 2 \cos (2 \pi \Phi m_1  + k_y) f_{m_1} \nonumber\\
&+ \left( 1 + \frac {t_1}{t_0}e^{i (2 \pi \Phi m_1 + k_y)} \right) f_{m_1+1}, \label{Difference Eq.}
\end{align}
with eigenvalues
\begin{equation}
\lambda = \frac{\epsilon^2 - (t^2_0 + 2 t^2_1)}{t_0 t_1}.
\end{equation}

This equation is periodic with $q$ and the Bloch condition then allows us to write
\begin{equation}
f_{m_1+q}(k_x)= e^{i q k_x} f_{m_1}(k_x).
\end{equation}
The Bloch condition reduces the calculation of the eigenvalues to diagonalization of a $q\times q$ matrix at every value of $k_x,k_y$. Because of the $q$-fold periodicity, unique values of $k_x$ are limited to the region $-\frac{\pi}{q}\leq k_x<\frac{\pi}{q}$ while $-\pi\leq k_y<\pi$ defining the magnetic Brillouin zone. Every value of $\lambda$ found at $k_x,k_y$ defines two $\epsilon$ symmetric around zero reflecting the bipartite symmetry. Here, $\lambda$ are the eigenvalues of
\begin{widetext}
\begin{equation}
\begin{pmatrix}
 2\cos\left(2\pi\frac{p}{q}0+k_y\right)& 1+\frac{t_1}{t_0}e^{i(2\pi\frac{p}{q}0+k_y)} & \cdots & \cdots & e^{-iqk_x}+\frac{t_1}{t_0}e^{i(2\pi\frac{p}{q}-k_y-qk_x)})\\\\
1+\frac{t_1}{t_0}e^{-i(2\pi\frac{p}{q}0+k_y)} & 2\cos\left(2\pi\frac{p}{q}1+k_y\right) &1+\frac{t_1}{t_0}e^{i(2\pi\frac{p}{q}1+k_y)}&  \cdots & \ldots \\\\
\vdots & \ddots &\ddots & \ddots & 1+\frac{t_1}{t_0}e^{i(2\pi\frac{p}{q}(q-2)+k_y)} \\\\
e^{iqk_x}+\frac{t_1}{t_0}e^{-i(2\pi\frac{p}{q}-k_y-qk_x)}& \cdots & \cdots & 1+\frac{t_1}{t_0}e^{-i(2\pi\frac{p}{q}(q-2)+k_y)} & 2\cos\left(2\pi\frac{p}{q}(q-1)+k_y\right)
\end{pmatrix}. \label{matrix}
\end{equation}
\end{widetext}
While it is possible to do numerical diagonalization at each point in the Brillouin zone, such a numerically costly approach is not necessary as long as one is interested in the width of the bands and gaps in the spectrum. The maxima and the minima of the bands appear at the same $k$-space points for all bands at a given flux. Determination of these special points enables us to plot the spectrum at any flux by diagonalizing two $q\times q$ matrices \cite{Hofstadter,Rammal,HofstadterWannier}.

\begin{table*}
\caption{ Critical values for extremum points of the energy bands for flux $\Phi=\frac{p}{q}$.} \label{crit. point table}
\begin{ruledtabular}
\begin{tabular}{c|c|c}
& $\left(\frac{t_1}{t_0}\right)^q>\frac{1}{2}$ &$\left(\frac{t_1}{t_0}\right)^q<\frac{1}{2}$\\ \hline
q even &
\begin{math}
(qk_x,qk_y)=\left(\pm\cos^{-1}\frac{1}{2}\left(\frac{t_1}{t_0}\right)^{-q}, \pm\cos^{-1}\frac{1}{2}\left(\frac{t_1}{t_0}\right)^{-q}\right)
\end{math}
&$(k_x,k_y)=(0,0)$ \\
&$(qk_x,qk_y)=(-\pi,\pi)$   &$(qk_x,qk_y)=(-\pi,\pi)$\\ \hline
q odd &$(qk_x,qk_y)=\left(-\pi\pm\cos^{-1}\frac{1}{2}\left(\frac{t_1}{t_0}\right)^{-q}, \pi\pm\cos^{-1}\frac{1}{2}\left(\frac{t_1}{t_0}\right)^{-q}\right)$ & $(qk_x,qk_y)=(-\pi,\pi)$ \\
&$(k_x,k_y)=(0,0)$   &$(k_x,k_y)=(0,0)$\\
\end{tabular}
\end{ruledtabular}
\end{table*}
The secular determinant of the eigenvalue equation has a simple dependence on $k_x,k_y$. Due to the $2\pi/q$ periodicity of the energy in both $k_x$ and $k_y$, there is only one momentum-dependent term, $\mathbb{D}(k_x,k_y)$, in the determinant. This term is independent of $\lambda$ and can be calculated as
\begin{align}
\mathbb{D}(k_x,k_y)&= (-1)^{q+1}\cos(q k_x)+(-1)^{p(q-1)}\cos(qk_y)\nonumber\\
&+(-1)^{p(q-1)+q+1} \left(\frac{t_1}{t_0}\right)^q \cos(q(k_x + k_y)).
\end{align}
The extrema of this expression define the points in the Brillouin zone at which all bands have their maximum or minimum energy. These critical $\vec{k}$ points can be found as given in Table.\ref{crit. point table}.

Both the calculation of the tight-binding parameters for the experimental optical lattice and the calculation of the Hofstadter butterfly for those tunneling rates are numerically efficient. This allows us to calculate the butterfly spectra in all the experimentally accessible regimes. In the next section, we focus on qualitatively different spectra that can be obtained and discuss the non-trivial transition between them.

\section{THE ENERGY SPECTRA}
The experiment in Ref.\cite{EsslingerDirac} demonstrates an extremely versatile optical lattice. While it is possible to calculate the butterfly spectrum for any set of experimental parameters, in this paper, we are mainly concerned with what kinds of topologically distinct spectra are accessible. The Hofstadter butterfly features an infinite number of gaps due to its fractal nature and closure of any one of these gaps creates a topologically different spectrum in principle. However, the smaller gaps in the spectrum are unlikely to be observed experimentally due to the finite size and temperature effects \cite{OnurHuiOktel}. The closing of the largest gaps in the spectra signals drastic changes in the behavior of the system and is more likely to have experimental signatures.

As characterized by their major gaps, the butterfly spectra for the adjustable lattice generally falls into one of the two categories. The spectrum is either similar to two square lattice Hofstadter butterflies separated by a large gap or it is similar to the butterfly of the honeycomb lattice (Hofstadter-Rammal butterfly). As we show below, the transition between these two general classes is not abrupt. The major gap near zero energy does not close at a single critical lattice parameter, but, this gap fractures into smaller gaps through gap-closings at principle fractions of flux. We discuss the physics of this process by identifying four representative points in the parameter space and displaying their spectra \cite{SupplyGif}.

The geometry of the adjustable lattice in Ref.\cite{EsslingerDirac} changes mainly by the reorganization of the potential inside a unit cell while the Bravais lattice always remains square. Hence, the different regimes such as "dimer", "checkerboard", etc. can be best understood by investigating the potential within a real-space unit cell. This potential can have one or two minimum points and the transition from one to two minima signals the change from the checkerboard to the dimer lattice. If the potentials are deep enough to form a tight-binding lattice, the Wannier functions localized around these minima have very small overlaps with the neighboring unit cell. The lowest two states form rather narrow bands which are separated by a gap controlled by the overlaps inside the unit cell. Thus, within the dimer and checkerboard regimes, the spectrum under the magnetic field consists of two square lattice Hofstadter butterflies.

\begin{figure}
\includegraphics[width=0.47\textwidth]{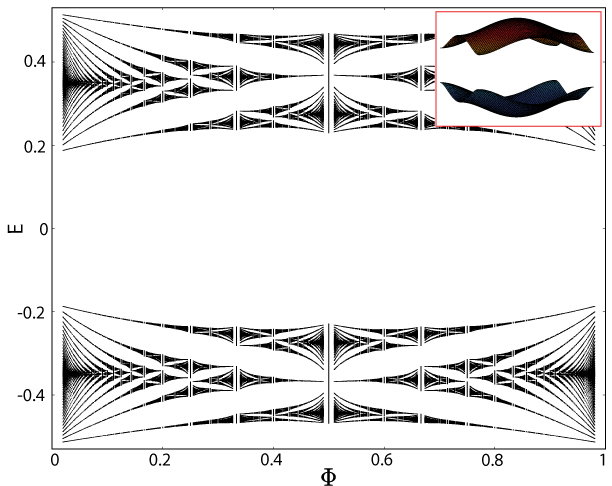}
\caption{(Color online) Energy spectrum as a function of flux per unit cell. For $V_{\bar{x}}=4.5,V_x=1,V_y=5$ and $\alpha=0.95$, creating a lattice in the Dimer regime. All energies are in units of $E_R$, flux is measured in units of flux quantum. Tunneling amplitudes are $t_0 = 0.3488$ and $t_1 = 0.0843$ in units of recoil energy. The spectrum is formed by two square lattice Hofstadter butterflies separated by a large gap controlled by $t_0$. The width of both butterflies are proportional to $t_1$. Inset is used to display energy bands as a function of $k_x,k_y$ at zero magnetic field. The gap at zero magnetic field is mostly retained at non-zero field. }
\label{Butterfly Vxbar4_5}
\end{figure}

As an example, in Fig.\ref{Butterfly Vxbar4_5} we display the results of our calculations for the experimental parameters $V_{\bar{x}}=4.5,V_x=1,V_y=5$ and $\alpha=0.95$. For these parameters, the real-space unit cell has two minima close to each other but far away from the minima in the neighboring cells. This arrangement was called the dimer lattice (see Fig.1 in Ref.\cite{EsslingerDirac}). We find the tunneling amplitudes $t_0 = 0.3488,\,t_1 = 0.0843$ for these parameters and calculate the corresponding butterfly. The spectrum is clearly formed by two square lattice Hofstadter butterflies separated by a large gap. This separation is controlled by the in-cell hopping parameter $t_0$. The upper (lower) butterfly is formed by the square lattice of anti-bonding (bonding) states in each unit cell.

The lattice gets closer to the honeycomb regime as the two minima within the unit cell move away from each other. The overlap between localized states at each minimum decreases with the increasing distance between them. Simultaneously, the distance between minima in adjacent cells decreases, increasing the corresponding tunneling rate. While the tight-binding amplitudes are not determined solely by the distance between the minima, this rough picture points to $t_0$ decreasing and $t_1$ increasing as honeycomb regime is approached. The zero field band structure shows an abrupt transition evidenced by the creation of Dirac points at the zone corners. As calculated above, the Dirac points exist for $\frac{t_1}{t_0}\geq\frac{1}{2}$ and this value was generally used to mark the boundary of the honeycomb regime \cite{EsslingerDirac,MontambauxPRL}.

\begin{figure}
\includegraphics[width=0.47\textwidth]{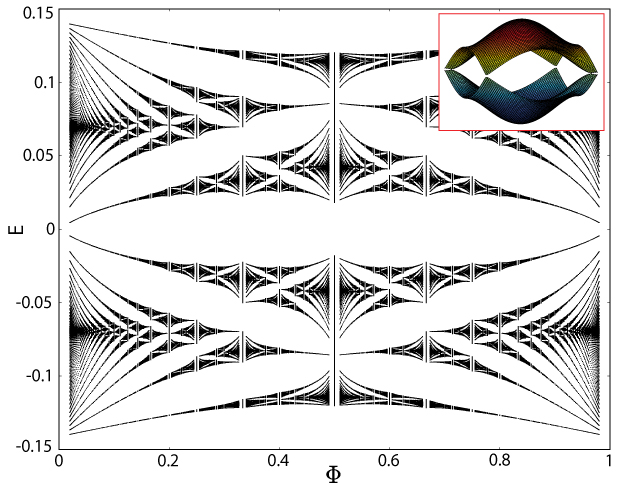}
\caption{(Color online) Hofstadter butterfly for $V_{\bar{x}}=12,V_x=0.79,V_y=6.45$ and $\alpha=0.9$, which is at the boundary of the honeycomb regime. Tunneling amplitudes are $t_0 = 0.0691$ and $t_1 = 0.0361$ in units of recoil energy. Comparing with Fig.\ref{Butterfly Vxbar4_5}, one can see that the two stacked Hofstadter butterflies get closer and touch only at the edges. As displayed in the inset, there are Dirac points at zero magnetic field band structure. However, for all non-zero magnetic fields, there is a finite gap at zero energy. }
\label{Butterfly Vxbar12}
\end{figure}

We calculate the spectrum at non-zero field for lattices which have Dirac points at zero field. We find that the existence of the Dirac points at zero field does not guarantee the existence of zero energy states at finite magnetic field. As a concrete example, in Fig.\ref{Butterfly Vxbar12} we plot the butterfly for $t_0 = 0.0691,\,t_1 = 0.0361$ which corresponds to experimental values $V_{\bar{x}}=12,V_x=0.79,V_y=6.45$ and $\alpha=0.9$. For these values, the major gaps of the spectrum resemble two square lattice butterflies touching only at their edges. The gap between these two segments lies unbroken for all flux values. Hence, although the zero field spectrum has Dirac points like the honeycomb lattice, the global shape of the spectrum under magnetic field is closer to two stacked square lattice butterflies.

As another example, consider a lattice deeper in the honeycomb regime. Tunneling amplitudes are $t_0 = 0.0441,\:t_1 = 0.0361$ which can be realized by $V_{\bar{x}}=13.5,V_x=0.79,V_y=6.45$ and $\alpha=0.9$. Again, although Dirac points exist at zero field, there are large gaps at zero energy. However, the overall central gap observed in the Fig.\ref{Butterfly Vxbar12} is now broken up into four parts. These closings happen at flux $p/q=1/2,1/3,2/3$ but nowhere else.
\begin{figure}
\includegraphics[width=0.47\textwidth]{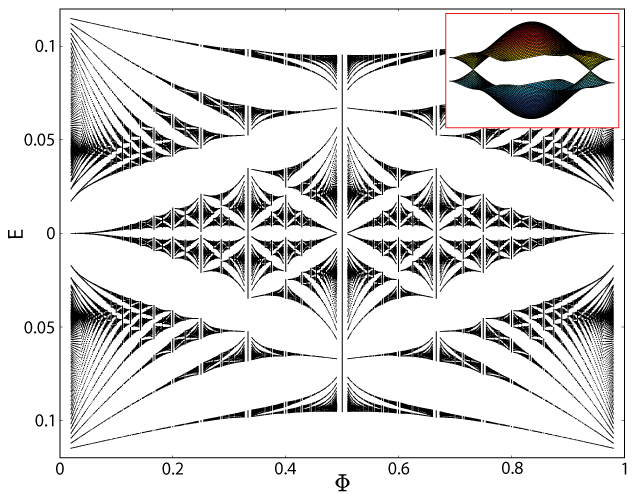}
\caption{(Color online) Hofstadter butterfly for $V_{\bar{x}}=13.5,V_x=0.79,V_y=6.45$ and $\alpha=0.9$, deeper in the honeycomb regime. Tunneling amplitudes are $t_0 = 0.0441$ and $t_1 = 0.0361$ in units of recoil energy $E_R$. The central gap is closed at the edges ($\Phi=0,1$) and the flux values $p/q = 1/2, 1/3, 2/3$ but nowhere else. For $\frac{t_1}{t_0} = 0.8186$, Eq.\ref{analytic closure} shows that all fluxes with $q\geq4$ have a gap at zero energy. The central gap is now split into four parts. Comparing the inset with the almost identical inset of Fig.\ref{Butterfly Vxbar12}, reveals the inability of zero magnetic field band structure to determine the spectrum for non-zero field. }
\label{Butterfly q23closed}
\end{figure}

The critical ratio $t_1/t_0$ for which the central gap closes at a particular flux can be calculated analytically. If the gap is closed, the Eq.\ref{S.E.A}, Eq.\ref{S.E.B} must admit a zero energy solution. At zero energy, these equations relate the coefficients of the wave function in adjacent unit cells. $q$ fold iteration connects two coefficients which are equivalent up to a phase due to the Bloch condition:
\begin{equation}
\left(\frac{t_1}{t_0}\right)^{-q}+(-1)^{p(q-1)}e^{ik_xq}=(-1)^qe^{ik_yq}.
\end{equation}
This relation is satisfied for some point in the Brillouin zone iff
\begin{equation}
\frac{t_1}{t_0}\geq\frac{1}{2^{1/q}}. \label{analytic closure}
\end{equation}
Thus, for any flux $\Phi=\frac{p}{q}$, there is no gap near zero energy if $t_1 > \frac{t_0}{2^{1/q}}$ regardless of the value of $p$. As the lattice evolves from the dimer to the honeycomb geometry, the first point where the central gap closes is at zero magnetic field for $t_1 =\frac{t_0}{2}$, which is the formation of Dirac points at zone corners. However, for $0.5 <  \frac{t_1}{t_0} <  \frac{1}{\sqrt{2}} \simeq 0.7071$, the gap at zero energy remains open at any non-zero magnetic field. The next flux for which the gap closes is $\Phi = \frac{1}{2}$, followed by $\Phi = \frac{1}{3}, \frac{2}{3}$ at $\frac{t_1}{t_0} = \frac{1}{2^{1/3}} \simeq 0.7937$. As an example, we plot the butterfly for $\frac{t_1}{t_0} = 0.8186$ in Fig.\ref{Butterfly q23closed} where one can observe the existence of the central gap except for three points discussed above.

\begin{figure}
\includegraphics[width=0.47\textwidth]{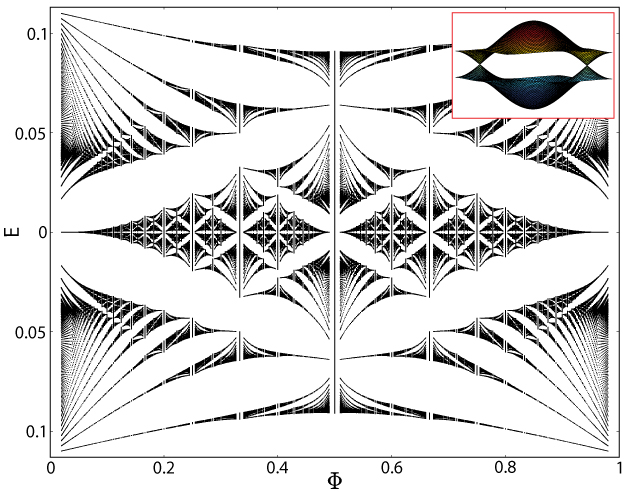}
\caption{(Color online) Hofstadter butterfly for $V_{\bar{x}}=14,V_x=0.79,V_y=6.45$ and $\alpha=0.9$, the parameters used to simulate graphene \cite{EsslingerGraphene}. Tunneling amplitudes are $t_0 = 0.0391$ and $t_1 = 0.0361$ in units of recoil energy. For these values, the spectrum nearly matches the ideal honeycomb butterfly Fig.\ref{graphene}, except for minor gaps. The central gap is open for the flux values with denominator $q\geq9$, producing narrow gaps near zero energy. As seen in the inset, the change in the zero field band structure is only due to the movement of Dirac points. }
\label{Butterfly Vxbar14}
\end{figure}

\begin{figure}
\includegraphics[width=0.47\textwidth]{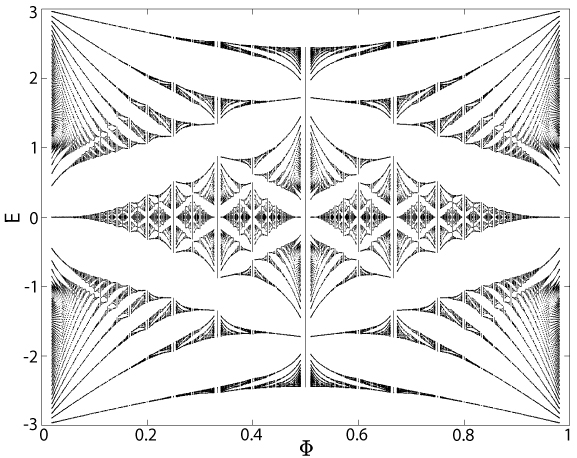}
\caption{ Hofstadter butterfly of the ideal honeycomb lattice ($t_0=t_1$) as obtained in Ref.\cite{Rammal}. Energy is plotted in units of $t_0$. Notice that there is no gap at zero energy for any value of flux in contrast to Fig.\ref{Butterfly Vxbar14}. Still, the overall structure is remarkably similar. }
\label{graphene}
\end{figure}

The stitching of two square lattice butterflies to form a honeycomb lattice butterfly is a topologically non-trivial process. As we calculated above, this evolution involves the creation and then subsequent closing of infinitely many gaps. Even more surprising is the fact that the critical values for gap-closings at zero energy follow a simple yet non-analytic function which depends only on the denominator of the flux. The vanishing of the central gap is completed only when $t_1 = t_0$, which identically corresponds to the honeycomb lattice. It is thus natural to ask whether the lattice geometry that was used to simulate graphene \cite{EsslingerGraphene} retains this similarity at finite field. In Fig.\ref{Butterfly Vxbar14}, we show the butterfly for the experimental parameters used to simulate graphene. For these parameters, we calculate that the central gap is closed for all flux values with denominators less than $q=9$. However, comparing this spectrum with the butterfly for an ideal honeycomb lattice Fig.\ref{graphene}, we see that while the gaps at zero energy have not completely closed, they are sufficiently small so that the overall shape of the butterfly is almost the same. As long as the energy resolution of a measurement is coarser than the size of these gaps, the same parameters can be used to simulate graphene under a magnetic field.

While the rough structure of the spectrum as characterized by the major gaps follows the physical picture given in this section, it is important to describe topological properties of the evolution in detail. In the next section, we calculate the Chern numbers for the gaps in the spectrum and show that bands exchange topological invariants through closing and subsequent opening of gaps. While infinitely many closings take place, all of them occur for minor gaps in the spectrum and are unlikely to be observed within current experimental energy resolution.

\section{CHERN NUMBERS}
The energy bands for a periodic potential under a magnetic field have a topological invariant called the first Chern number which is directly related to the contribution of that band to Hall conductivity \cite{TKNN}. As long as a band is isolated by gaps in energy, this integer cannot change due to changes in the lattice potential or the magnetic field. Thus, integer Chern numbers can be used to label the gaps in the butterfly spectra \cite{Avron}. If the Fermi energy lies in a gap labeled with Chern number $\nu$, the Hall conductivity for a non-interacting system of fermions is $\sigma_{xy}=\nu e^2/h$.

Since Chern numbers are topologically protected against small perturbations and directly related to measurable quantities, they have attracted both theoretical and experimental interest recently \cite{BlochChern,GoldmanChern,DuanChern,KetterleQHE,CooperChern}. A number of methods for Chern number and Berry curvature measurement in cold atom systems have been proposed and implemented. A method which is not only useful for experimental Chern number measurement but also for numerical calculation relies on the St\v{r}eda formula \cite{Streda}. St\v{r}eda formula connects the response of the density to changes in magnetic flux with the Hall conductivity, hence the Chern number.
\begin{eqnarray*}
\sigma_{xy} = \frac{\partial N} {\partial B},
\end{eqnarray*}
where $N$ is the number of levels lying below the Fermi energy. As density probes are more common in cold atom experiments compared to transport measurements, this relation is exceptionally suited for ultracold systems \cite{OnurHuiOktel}. Adapting this formula to the system in consideration, the Chern number for any gap is
\begin{equation}
\nu=\frac{\partial n}{\partial \Phi},
\end{equation}
where $n$ is the density per unit cell for fermions with Fermi energy lying in that gap. Numerically we calculate Chern numbers by implementing the derivative by a finite difference. The calculation is reduced to counting of the eigenvalues of the matrix Eq.\ref{matrix} below the Fermi energy for two closely spaced flux values \cite{OnurHuiOktel}.
\begin{figure}
\includegraphics[width=0.47\textwidth]{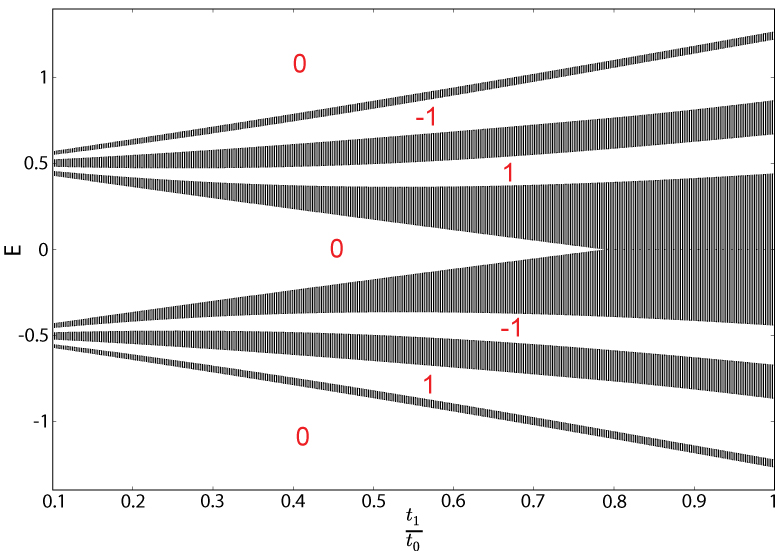}
\caption{(Color online) The evolution of bands from the checkerboard to the honeycomb regime as a function of $\frac{t_1}{t_0}$ for flux $\Phi=1/3$. Gaps are labeled with their Chern number. The gaps with non-zero Chern numbers are open for all $0<\frac{t_1}{t_0}\leq1$. The central gap closes at $\frac{t_1}{t_0}=\frac{1}{2^{1/3}}$. Near $\frac{t_1}{t_0}=0.1$, the gaps are the gaps of the two stacked square lattice butterflies. In the other limit, they connect to the largest gaps of the honeycomb lattice without gap closings. }
\label{q1/3 Fig}
\end{figure}

We first test the observation made in the previous section that the largest gaps except the central gap in the spectrum remain open during the evolution from the checkerboard to the honeycomb lattice. As can be seen in Fig.\ref{q1/3 Fig} calculated for flux $\Phi=1/3$, the gaps with Chern numbers $\pm1$ remain open for all $0<\frac{t_1}{t_0}\leq1$. In one limit, these gaps can be interpreted as the major gaps of two stacked square lattice butterflies while in the other limit they are the largest gaps of the honeycomb lattice. The central gap with Chern number zero closes at $\frac{t_1}{t_0}=\frac{1}{2^{1/3}}$.

The simple evolution for the largest gaps is not repeated by smaller gaps. In general, many small gaps are closed and reopened as $\frac{t_1}{t_0}$ increases \cite{SupplyGif}. The number of bands at flux $\Phi=p/q$ is $2q$ for two stacked  square lattices and $2q-1$ for the honeycomb lattice. Apart from the closing of the central gap, the number of gap closures must be equal to the number of gap openings. However, two bands can change their Chern numbers through touching and splitting, conserving the total Chern number.
\begin{figure}
\includegraphics[width=0.47\textwidth]{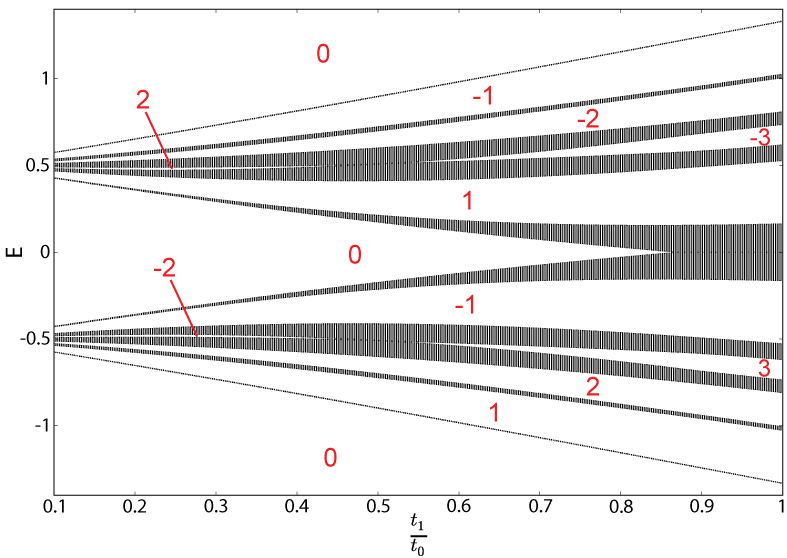}
\caption{(Color online) The evolution of bands from the checkerboard to the honeycomb at $\Phi =1/5$ as a function of $\frac{t_1}{t_0}$. Gaps are labeled with their Chern number. The sequence of the Chern numbers in the spectrum near $\frac{t_1}{t_0}=0.1$ is 1, 2, -2, -1, 0, 1, 2, -2, -1 as expected for two stacked square lattice butterflies. In the other limit, the sequence is 1, 2, 3, -1, 0, 1, -3, -2, -1 valid for the honeycomb lattice. The change in the sequence is possible due to the third (eighth) and the fourth (seventh) bands touching and exchanging Chern numbers at $\frac{t_1}{t_0} = 0.5$. The central gap closes at $\frac{t_1}{t_0} = \frac{1}{2^{1/5}}$. }
\label{q1/5 Fig}
\end{figure}

As an example, we display the bands at $\Phi = \frac{1}{5}$ as a function of $\frac{t_1}{t_0}$ in Fig.\ref{q1/5 Fig}. Near $\frac{t_1}{t_0}\simeq 0$ , the gaps in the spectrum have the Chern numbers 1, 2, -2, -1, 0, 1, 2, -2, -1 in the order of increasing energy. This sequence is replaced by 1, 2, 3, -1, 0, 1, -3, -2, -1 in the honeycomb lattice limit $\frac{t_1}{t_0} = 1$. We see that in addition to the central gap closing at $\frac{t_1}{t_0} = \frac{1}{2^{1/5}}$, the third (eighth) and the fourth (seventh) bands touch and exchange Chern numbers at $\frac{t_1}{t_0} = 0.5$.

\begin{figure}
\includegraphics[width=0.47\textwidth]{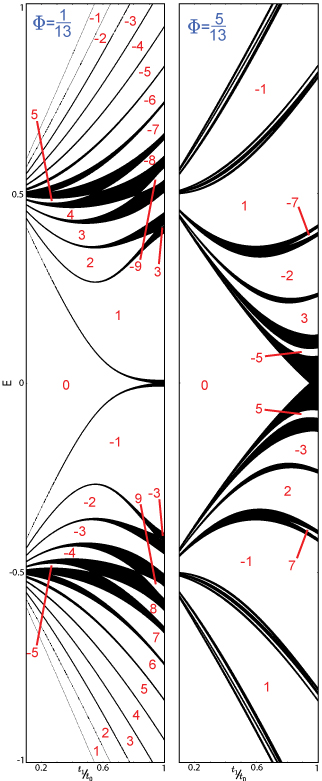}
\caption{(Color online) Evolution of the bands and gap Chern number for $\Phi=1/13,\,\Phi=5/13$. For large denominators, the evolution contains many band crossings and Chern number transfers. These transfers are necessary for the sequence of gap Chern numbers to connect the two limits. Notice that the largest gaps remain open throughout. Chern numbers for the smallest gaps are not displayed for visual clarity. }
\label{long fig}
\end{figure}

In general, for larger denominator $q$, there will be more contact between the bands throughout the evolution Fig.\ref{long fig}. As the number of bands for the honeycomb lattice is less than the number of bands for two stacked square lattice butterflies by just one, all the gaps except the central gap are related by a finite number of Chern number transfers between the bands. In that sense, it is not surprising that the Diophantine equation which constraints the gap Chern number is the same for square and honeycomb lattices \cite{Diophantine}.

In particular, the experimental setup used in Ref.\cite{EsslingerGraphene} to mimic graphene will differ from graphene under a magnetic field for flux values with denominators greater than $q =8$. The largest gap that is open for the optical lattice but not for the ideal honeycomb lattice has an energy width which is almost one percent of the full band width. Unless the experiment probes with this energy resolution, the butterflies of the optical lattice and graphene are equivalent.

As the experiment can access all $\frac{t_1}{t_0}$ values, it is possible to observe adiabatic Chern number transfer between the bands. Consequently, a lattice with large enough gaps which are adiabatically connected to honeycomb lattice gaps can be manufactured. A measurement of the integer Chern numbers within the largest gap sequence near zero energy is possible with the experimental parameters used in Ref.\cite{EsslingerGraphene}. Such a measurement would amount to a demonstration of the integer quantum Hall effect for Dirac fermions \cite{QHEgraphene} in cold atom systems.

\section{CONCLUSION}
Realization of artificial magnetic fields in optical lattices enables the observation of Hofstadter butterfly spectrum with cold atoms. This self-similar energy spectrum is highly specific to the lattice geometry. Hence, the latest advances in creating adjustable lattice geometries provide the opportunity to study the dependance of the Hofstadter butterfly to the lattice parameters.

We calculated the energy spectra for experimentally realized potentials \cite{EsslingerDirac} at non-zero magnetic field. The effect of the magnetic field is taken into account by using Peierls substitution on a nearest neighbor tight-binding model for the potential.

Our results indicate that if an artificial magnetic field is created in the tunable lattice used in Ref.\cite{EsslingerDirac,EsslingerGraphene}, it would be possible to observe the evolution of the spectrum from two square lattice butterflies to a honeycomb lattice butterfly. As the largest gaps remain open throughout the transition, Chern numbers measured at any point within these gaps are equivalent to the ideal honeycomb lattice spectrum. The most prominent gap that closes during the transition is the central gap, which vanishes by sequential creation of Dirac points in the spectrum at rational flux values. We calculated these critical points for any rational flux $\Phi=\frac{p}{q}$ in terms of the tight-binding parameters $\frac{t_1}{t_0}=\frac{1}{2^{1/q}}$.

Our most important conclusion is that the existence of Dirac points at zero magnetic field does not guarantee that the Hofstadter butterfly is similar to the spectrum of the ideal honeycomb lattice. As the $\frac{t_1}{t_0}$ ratio is increased, the general structure of the butterfly spectrum approximates the honeycomb butterfly better. The two butterflies are identical only when $\frac{t_1}{t_0}=1$. For any ratio smaller than one, infinitely many but extremely narrow gaps are present. Bands in the spectrum exchange Chern numbers through the closing and reopening of these minor gaps.

In principle, the evolution of this lattice makes it possible to observe Chern number transfer between the adjacent bands through band touching. However, we found that the gaps which close and reopen are quite narrow compared to the full band width. The energy resolution to investigate this process can only be obtained by using large lattices so that finite size effects do not wash out these small gaps. If these small gaps are not resolved, the artificial graphene created by the Zurich group would successfully simulate the Hofstadter butterfly of graphene.

\begin{acknowledgments}

F.Y. is supported by T\"{u}rkiye Bilimsel ve Teknolojik Ara\d{s}t{\i}rma Kurumu (T\"{U}B\.{I}TAK) Scholarship No. 2210 and F.N.\"{U}. is supported by T\"{U}B\.{I}TAK Scholarship No.2211. This work is supported by T\"{U}B\.{I}TAK Grant No. 112T974.

\end{acknowledgments}

\bibliography{PaperPRA}

\end{document}